\pdfoutput=1
\documentclass[11pt]{article}
\usepackage[margin=1in]{geometry}
\usepackage{amsmath,amssymb}
\usepackage{amsthm}
\usepackage{graphicx}
\usepackage{float}
\usepackage{hyperref}

\newtheorem{proposition}{Proposition}

\title{On the Connection Between Differential Population Growth Rate and Epidemic Reproduction Numbers}
\author{Hong Qin\textsuperscript{*} \\[4pt]
{\normalsize School of Data Science, Department of Computer Science,} \\
{\normalsize Old Dominion University, Norfolk, VA 23529, USA} \\[4pt]
{\normalsize ORCID: 0000-0001-9007-4622}}
\date{}

\begin{document}

\maketitle

\bigskip

\begin{abstract}
During pandemics, public health agencies need to rapidly assess whether a new
viral variant is more transmissible than existing ones. When two viral variants
co-circulate, their relative fitness can be quantified in several related ways:
as a selective coefficient in the tradition of Kimura's population genetics, as
the differential population growth rate (DPGR) estimated from genomic
surveillance data, or as a contrast in epidemic reproduction numbers $R_t$. We
show that DPGR estimates a pairwise growth-rate difference. Under a specified
generation-interval model, this growth-rate difference can be transformed into
a contrast in reproduction-number space; in the equal-generation-time SIR
special case, the relationship reduces to a scaled difference in
variant-specific $R_t$. Closely related growth-rate contrasts also appear
inside multinomial logistic (softmax) and growth-advantage random walk (GARW)
frameworks, although those methods differ from DPGR in likelihood structure,
smoothing, priors, and data inputs. We evaluate the theory empirically across
five analyses spanning SARS-CoV-2 (three analyses, $r=0.56$ to $0.78$) and
influenza H1N1 and H3N2 (two analyses, $r=0.31$ to $0.38$), totaling over
2{,}200 matched data points, with SIR simulation confirming recovery of the
expected mapping (slope $= 0.99$) when the true $R_t$ is known. Applied
retrospectively to five major SARS-CoV-2 variant transitions, sustained DPGR
signals detected emerging fitness advantages 43 to 65~days before variant
dominance, with 95\% sign accuracy in our analysis. DPGR is approximately
transitive across 115{,}624 variant triplets, consistent with an additive
lineage-level growth-rate representation, approximately zero for selected
functionally similar sub-lineages, and directionally consistent across
11 countries. The core algebraic statements are formalized in Lean~4 with
Mathlib. These results connect sequence-count-based variant fitness estimates
to reproduction-number contrasts through an assumption-explicit growth-rate
bridge.
\end{abstract}

\medskip


\bigskip

\section{Introduction}

When a new viral variant is detected during a pandemic, public health agencies
face an urgent question: will this variant outcompete existing ones, and if so,
how quickly? Answering this currently requires fitting complex transmission
models to absolute incidence data, estimating generation-time distributions,
and waiting for sufficient case counts to accumulate. Yet the most rapidly
available signal is often the simplest: the relative abundance of variant
sequences in routine genomic surveillance. If the mathematical relationship
between sequence-derived fitness measures and epidemic reproduction numbers
were made explicit, variant risk assessment could begin as soon as sequence
data appear, weeks before traditional epidemiological estimates become
reliable.

The question of how to measure the relative fitness of competing biological
types has a long history. In classical population genetics, Kimura showed that
under deterministic selection in a large population, the log-ratio of two
allele frequencies changes linearly in time at a rate equal to their selective
coefficient difference (Kimura 1962). His neutral theory further
established that $s=0$ (no systematic trend in the log-ratio) is the null
hypothesis against which selection is tested (Kimura 1968).

Six decades later, Pantho et al.\ (2025) independently arrived at the same
mathematical structure in the context of viral genomic surveillance. Their
differential population growth rate (DPGR) estimates the slope of
$\log_{10}(N_i/N_j)$ over a sliding time window, which is the selective
coefficient of classical theory, applied to co-circulating viral lineages
(Pantho et al. 2025).\footnote{In their application, time is measured in days and
the regression is performed on a sliding time window chosen so that the
log-ratio is approximately linear.}
The DPGR framework has since been extended to influenza strain
dominance across regions and seasons (Uddin et al. 2025), to predicting
variant fitness from full viral genome sequences (Annan et al. 2025), and
to genome-wide association analysis of fitness-linked mutations
(Hatami et al. 2026).

Meanwhile, several groups have developed methods to estimate variant-specific
fitness from sequence frequencies: Kimura et al.'s Bayesian multinomial
logistic (softmax) model (Kimura et al. 2022), Figgins and Bedford's
growth-advantage random walk (GARW) (Figgins and Bedford 2025), and Obermeyer
et al.'s PyR0 framework (Obermeyer et al. 2022). Each estimates a
variant-specific growth advantage and converts it to a relative effective
reproduction number $R_e$ by assuming a generation-time distribution.

A natural question arises: how are these quantities (the classical selective
coefficient, DPGR, and contrasts in epidemic reproduction numbers) related?
We show that DPGR estimates the pairwise growth-rate contrast. Under a
short-window two-variant SIR approximation, and conditional on shared
susceptibility assumptions, DPGR maps to a scaled difference in
variant-specific $R_t$ (Proposition~\ref{prop:sir-bridge}). Under a
fixed-generation-time model, it maps to a scaled log-ratio of $R_e$ values.
More generally, for any monotone growth-rate-to-reproduction-number map
$\Phi$, DPGR induces a pairwise contrast only after the baseline
growth rate and the map $\Phi$ have been specified
(Proposition~\ref{prop:abstract-bridge}). Related growth-rate contrasts also
appear inside softmax and GARW parameterizations. The core algebraic results
are formalized in Lean~4 with Mathlib (Appendix), and evaluated empirically
across five analyses spanning SARS-CoV-2 and influenza.

Throughout, $R_t$ denotes the time-varying effective reproduction number in an
SIR model, and $R_e$ denotes the growth-rate-derived reproduction number via
the Wallinga-Lipsitch mapping (Wallinga and Lipsitch 2007).

\section{Population Genetics Foundation}

The connection between DPGR and classical fitness theory is immediate. Consider
two types $i$ and $j$ with fitnesses $w_i = 1+s_i$ and $w_j = 1+s_j$ in a
large population under deterministic selection. After $t$ generations, the
frequency ratio evolves as
\[
\frac{p_i(t)}{p_j(t)}
= \left(\frac{w_i}{w_j}\right)^t \frac{p_i(0)}{p_j(0)},
\]
so
\begin{equation}
\ln\!\left(\frac{p_i(t)}{p_j(t)}\right)
= t\,\ln\!\left(\frac{w_i}{w_j}\right) + C
\approx (s_i - s_j)\,t + C,
\label{eq:kimura_logratio}
\end{equation}
where the approximation holds for small $s$. This is the classical result: the
log-frequency ratio changes linearly with slope equal to the selective
coefficient difference (Kimura 1962).

Comparing \eqref{eq:kimura_logratio} with Pantho et al.'s definition
(Pantho et al. 2025),
\[
\log_{10}\!\left(\frac{N_i(t)}{N_j(t)}\right)
= \mathrm{DPGR}_{i,j}\,t + C,
\]
we obtain
\begin{equation}
s_i - s_j = \ln(10)\,\mathrm{DPGR}_{i,j}.
\label{eq:dpgr_selective_coeff}
\end{equation}
Thus DPGR is the classical selective coefficient, rescaled from natural to
base-10 logarithms.

One important difference separates the two settings. Kimura's model assumes
\emph{constant} fitness: the selective coefficient $s$ does not change over
time. In an epidemic, however, susceptible depletion causes the effective
fitness of each variant to change as immunity builds. The short-window
approximation used by Pantho et al.\ recovers the Kimura regime: within each
window, $S(t)/N$ changes slowly enough that the fitness difference is
approximately constant, and the log-ratio is approximately linear. The
time-varying $R_t$ tracks how this fitness evolves across windows.

From Kimura's neutral theory (Kimura 1968), the null hypothesis is
$s_i = s_j$, i.e., $\mathrm{DPGR}_{i,j} = 0$. A statistically significant
non-zero DPGR is therefore evidence of differential selection between the two
lineages.

\section{DPGR as a Difference in Growth Rates}

Pantho et al.\ define DPGR through the pairwise log-ratio model
\begin{equation}
\log_{10}\!\left(\frac{N_i(t)}{N_j(t)}\right)
= \mathrm{DPGR}_{i,j}\, t + C,
\label{eq:dpgr_def}
\end{equation}
where $N_i(t)$ and $N_j(t)$ are the two variant sub-populations in the chosen
time window and $C$ is a constant that absorbs lag effects and initial
conditions.

Equation \eqref{eq:dpgr_def} is written in base-10 logarithms because that is
the convention used in the DPGR paper. For epidemic modeling, it is often more
convenient to work with natural logarithms and Malthusian growth rates. Writing
\begin{equation}
N_k(t) \approx A_k e^{r_k t}, \qquad k\in\{i,j\},
\label{eq:exp_growth}
\end{equation}
gives
\begin{equation}
\ln\!\left(\frac{N_i(t)}{N_j(t)}\right)
= (r_i-r_j)t + C',
\label{eq:ln_ratio}
\end{equation}
so that
\begin{equation}
r_i-r_j = \ln(10)\,\mathrm{DPGR}_{i,j}.
\label{eq:dpgr_to_r}
\end{equation}
Thus, up to the factor $\ln(10)$, DPGR is a \emph{difference of
growth rates}. If one instead fits the log-ratio using natural logarithms, the
conversion factor disappears.

\section{Two-Variant SIR Derivation}

Consider a two-variant SIR-type model in a population of size $N$:
\begin{align}
\frac{dS}{dt} &= -\frac{S}{N}\bigl(\beta_i I_i + \beta_j I_j\bigr), \\
\frac{dI_i}{dt} &= \left(\beta_i\frac{S}{N} - \gamma_i\right) I_i, \\
\frac{dI_j}{dt} &= \left(\beta_j\frac{S}{N} - \gamma_j\right) I_j.
\end{align}
Here $\beta_k$ is the transmission rate and $\gamma_k$ is the removal or
recovery rate for variant $k$.

Over a sufficiently short window, if $S(t)$ changes slowly compared with the
window length, each infected sub-population is approximately exponential:
\begin{equation}
I_k(t) \approx I_k(t_0)\exp\!\left(r_k(t_0)(t-t_0)\right),
\end{equation}
with local growth rate
\begin{equation}
r_k(t) = \beta_k\frac{S(t)}{N} - \gamma_k.
\label{eq:r_sir}
\end{equation}

Define the variant-specific effective reproduction number by
\begin{equation}
R_{t,k} = \frac{\beta_k S(t)}{\gamma_k N}.
\label{eq:rt_def}
\end{equation}
This definition has a direct epidemiological meaning:
\begin{itemize}
    \item The factor $\beta_k S(t)/N$ is the instantaneous rate at which one
    infectious individual of variant $k$ generates new infections at time $t$,
    after accounting for the fact that only the fraction $S(t)/N$ of the
    population remains susceptible.
    \item In the SIR model, $1/\gamma_k$ is the mean infectious period for
    variant $k$.
    \item Multiplying these two quantities gives the expected number of
    secondary infections caused by one infectious individual of variant $k$ at
    time $t$, which is $R_{t,k}$.
\end{itemize}
Equivalently,
\[
\beta_k\frac{S(t)}{N} = \gamma_k R_{t,k}.
\]
Substituting this identity into \eqref{eq:r_sir}, we obtain the growth-rate
relation step by step:
\begin{align}
r_k(t)
&= \beta_k\frac{S(t)}{N} - \gamma_k \notag\\
&= \gamma_k R_{t,k} - \gamma_k \notag\\
&= \gamma_k\bigl(R_{t,k} - 1\bigr).
\label{eq:r_rt}
\end{align}
Thus $R_{t,k}>1$ corresponds to local exponential growth, $R_{t,k}<1$
corresponds to decline, and $R_{t,k}=1$ marks the threshold between the two.

\begin{proposition}[SIR bridge from DPGR to $R_t$ and $R_0$]
\label{prop:sir-bridge}
Under the two-variant SIR approximation,
\[
\mathrm{DPGR}_{i,j}(t)
= \frac{1}{\ln(10)}
\left[
\gamma_i\bigl(R_{t,i}(t)-1\bigr)
- \gamma_j\bigl(R_{t,j}(t)-1\bigr)
\right].
\]
If moreover $S(t)=N$, so that $R_{0,k}=\beta_k/\gamma_k$, then
\[
\mathrm{DPGR}_{i,j}(t)
= \frac{1}{\ln(10)}
\left[
\gamma_i\bigl(R_{0,i}-1\bigr)
- \gamma_j\bigl(R_{0,j}-1\bigr)
\right].
\]
\end{proposition}

Substituting \eqref{eq:r_rt} into \eqref{eq:dpgr_to_r} yields the main
connection recorded in Proposition~\ref{prop:sir-bridge}:
\begin{equation}
\mathrm{DPGR}_{i,j}(t)
= \frac{1}{\ln(10)}
\left[
\gamma_i\bigl(R_{t,i}-1\bigr)
- \gamma_j\bigl(R_{t,j}-1\bigr)
\right].
\label{eq:main_rt_connection}
\end{equation}

Equation \eqref{eq:main_rt_connection} shows that DPGR is not itself a
reproduction number. Rather, it is the \emph{difference in transmission growth
advantages} after translating reproduction numbers into exponential growth-rate
space.

\section{Important Special Cases}

\subsection*{Common infectious period or generation time}

If the two variants have approximately the same removal rate,
\[
\gamma_i \approx \gamma_j \approx \gamma,
\]
then \eqref{eq:main_rt_connection} simplifies to
\begin{equation}
\mathrm{DPGR}_{i,j}(t)
\approx \frac{\gamma}{\ln(10)}\bigl(R_{t,i} - R_{t,j}\bigr).
\label{eq:equal_gamma}
\end{equation}
With mean infectious period or generation time $T_g \approx 1/\gamma$, this can
also be written as
\begin{equation}
\mathrm{DPGR}_{i,j}(t)
\approx \frac{R_{t,i} - R_{t,j}}{T_g \ln(10)}.
\label{eq:tg_form}
\end{equation}
Therefore, under equal generation-time assumptions, DPGR is proportional to the
\emph{difference} in the variants' effective reproduction numbers.

\subsection*{Early epidemic or nearly fully susceptible phase}

If susceptible depletion is negligible so that $S(t)\approx N$, then
\[
R_{t,k} \approx R_{0,k} = \frac{\beta_k}{\gamma_k},
\]
and \eqref{eq:main_rt_connection} reduces to
\begin{equation}
\mathrm{DPGR}_{i,j}
\approx \frac{1}{\ln(10)}
\left[
\gamma_i\bigl(R_{0,i}-1\bigr)
- \gamma_j\bigl(R_{0,j}-1\bigr)
\right].
\label{eq:r0_general}
\end{equation}
If, in addition, $\gamma_i\approx\gamma_j\approx\gamma$, then
\begin{equation}
\mathrm{DPGR}_{i,j}
\approx \frac{\gamma}{\ln(10)}\bigl(R_{0,i} - R_{0,j}\bigr).
\label{eq:r0_equal_gamma}
\end{equation}
So in the early susceptible phase, DPGR is proportional to the difference in
basic reproduction numbers.

\section{Principled Interpretation}

The main conceptual distinction is:
\begin{itemize}
    \item DPGR is a \emph{relative} measure. It compares two variants inside the
    same epidemiological and surveillance setting by using one variant as an
    internal control for the other.
    \item $R_t$ is an \emph{absolute} measure. It describes how many secondary
    infections one infectious individual generates on average at time $t$.
    \item In an SIR approximation, DPGR lives in \emph{growth-rate
    space}. To convert it into $R_t$ space, one needs assumptions about
    generation time or recovery rate.
\end{itemize}

This distinction also explains why DPGR can be robust to some multiplicative
sampling biases: if both variants are affected by the same reporting or sampling
factor in the same window, that factor largely cancels in the ratio
$N_i/N_j$. By contrast, estimating $R_t$ typically requires absolute incidence
information and a generation-interval model.

\section{Relation Beyond the Strict SIR Assumption}

The SIR model implies an exponential infectious-period distribution. Under that
assumption, Wallinga and Lipsitch showed that the relation between growth rate
$r$ and reproduction number is linear:
\begin{equation}
R = 1 + \frac{r}{\gamma}.
\label{eq:wallinga_sir}
\end{equation}
More generally, the mapping from growth rate to reproduction number depends on
the generation-interval distribution.

\subsection*{Fixed generation time and a log-ratio representation}

If one approximates the generation interval by a point mass at its mean
$T_g$, then the Euler-Lotka relation gives the familiar approximation
(Wallinga and Lipsitch 2007)
\begin{equation}
R_{e,k}(t) \approx \exp\!\bigl(r_k(t) T_g\bigr),
\label{eq:re_exp}
\end{equation}
or equivalently
\begin{equation}
r_k(t) \approx \frac{\ln R_{e,k}(t)}{T_g}.
\label{eq:r_from_re}
\end{equation}
Here $R_{e,k}(t)$ denotes the effective reproduction number for variant $k$,
written with the notation $R_e$ to emphasize that this formula is a
growth-rate-to-reproduction-number mapping rather than a specifically
SIR-derived identity.

Substituting \eqref{eq:r_from_re} into
\eqref{eq:dpgr_to_r} yields
\begin{align}
\mathrm{DPGR}_{i,j}(t)
&\approx \frac{1}{T_g \ln(10)}
\left[
\ln R_{e,i}(t) - \ln R_{e,j}(t)
\right] \notag\\
&= \frac{1}{T_g}
\log_{10}\!\left(\frac{R_{e,i}(t)}{R_{e,j}(t)}\right).
\label{eq:dpgr_re_logratio}
\end{align}
So under a shared fixed-generation-time approximation, DPGR is a scaled
\emph{log-ratio} of the variants' effective reproduction numbers. This is
slightly different from the SIR formula in \eqref{eq:tg_form}, which expresses
DPGR as an approximate \emph{difference} in $R_t$ values.

The two forms agree to first order when the effective reproduction numbers are
near one. Using the linearization $\ln R_e \approx R_e - 1$ gives
\begin{equation}
\mathrm{DPGR}_{i,j}(t)
\approx \frac{R_{e,i}(t) - R_{e,j}(t)}{T_g \ln(10)},
\label{eq:dpgr_re_linearized}
\end{equation}
which recovers \eqref{eq:tg_form} when $R_e$ and $R_t$ are viewed as two
notations for the same time-varying effective reproduction number under the
shared-generation-time assumption.

Therefore, the most robust general statement is:
\begin{quote}
DPGR estimates a pairwise growth-rate advantage. It becomes a difference in
$R_t$, a log-ratio of $R_e$, or a difference in $R_0$ only after a
generation-time model is specified.
\end{quote}

This viewpoint admits a convenient abstract formulation, summarized in
Proposition~\ref{prop:abstract-bridge}.

\begin{proposition}[Abstract DPGR bridge]
\label{prop:abstract-bridge}
Suppose that a chosen generation-interval model induces a map $\Phi$ from local
growth rate to effective reproduction number, so that
\[
R_{t,k}(t)=\Phi\bigl(r_k(t)\bigr).
\]
Then DPGR still carries the information about the pairwise
growth-rate contrast,
\[
r_i(t)-r_j(t)=\ln(10)\,\mathrm{DPGR}_{i,j}(t),
\]
and, conditional on the baseline growth rate $r_j(t)$ and the chosen map
$\Phi$, induces the following pairwise contrast in reproduction-number space:
\[
R_{t,i}(t)-R_{t,j}(t)
= \Phi\bigl(r_j(t)+\ln(10)\,\mathrm{DPGR}_{i,j}(t)\bigr)-\Phi\bigl(r_j(t)\bigr).
\]
If, in the exponential-generation-time SIR case, $\Phi(r)=1+r/\gamma$, then
\[
\mathrm{DPGR}_{i,j}(t)
= \frac{\gamma}{\ln(10)}\bigl(R_{t,i}(t)-R_{t,j}(t)\bigr),
\]
provided the two variants share a common removal rate $\gamma$.
\end{proposition}

This is the principled bridge between the DPGR framework and classical epidemic
quantities.

\section{Connection to Sequence-Frequency and Base-Variant Models}

\subsection*{Kimura et al.'s Softmax-Based Method}

A related formulation appears in Kimura et al.\ (2022),\footnote{I.~Kimura et al.\ (2022), not to be confused with M.~Kimura's classical population genetics work (Kimura 1962; Kimura 1968), which provides the conceptual foundation discussed earlier.}
who model lineage replacement with a Bayesian multinomial logistic regression.
For lineages $l=1,\dots,m$, they define linear predictors
\begin{equation}
m_{l t} = a_l + b_l t,
\end{equation}
convert them to lineage frequencies through the softmax map
\begin{equation}
q_{l t} = \frac{\exp(m_{l t})}{\sum_{k=1}^m \exp(m_{k t})},
\end{equation}
and then model the observed lineage counts conditionally on the daily total
with a multinomial likelihood.

Although this is written as a multi-lineage softmax model, the pairwise
quantity of interest is immediate because the common denominator cancels:
\begin{equation}
\log\!\left(\frac{q_{i t}}{q_{j t}}\right)
= (a_i-a_j) + (b_i-b_j)t.
\label{eq:kimura_pairwise}
\end{equation}
Thus the Kimura slope difference $b_i-b_j$ is the pairwise
log-frequency growth advantage of lineage $i$ over lineage $j$.

This is the same estimand that DPGR targets. If the observed sequence counts
are proportional to the underlying lineage abundances within a short window,
then comparing \eqref{eq:kimura_pairwise} with \eqref{eq:ln_ratio} shows that
\begin{equation}
\ln(10)\,\mathrm{DPGR}_{i,j} = b_i-b_j.
\label{eq:dpgr_kimura_slope}
\end{equation}
So DPGR can be viewed as the pairwise slope implied by the softmax model,
written on a base-10 rather than a natural-log scale.

Kimura et al.\ then map the slope parameter to a lineage-specific relative
effective reproduction number using a fixed mean generation time $g$. To avoid
confusing this quantity with the growth rate $r_k$ already used above, let
\[
\rho_l = \exp(g b_l)
\]
denote the relative effective reproduction number of lineage $l$ with respect
to the chosen reference lineage. Pairwise,
\begin{equation}
\frac{\rho_i}{\rho_j} = \exp\!\bigl(g(b_i-b_j)\bigr).
\label{eq:kimura_pairwise_re}
\end{equation}
Combining \eqref{eq:dpgr_kimura_slope} and \eqref{eq:kimura_pairwise_re}
yields
\begin{align}
\mathrm{DPGR}_{i,j}
&= \frac{1}{g\ln(10)}
\ln\!\left(\frac{\rho_i}{\rho_j}\right) \notag\\
&= \frac{1}{g}
\log_{10}\!\left(\frac{\rho_i}{\rho_j}\right).
\label{eq:dpgr_kimura_re}
\end{align}

Therefore Pantho's DPGR and Kimura et al.'s softmax-based analysis share a
pairwise growth-rate contrast under local exponential/logistic growth
assumptions, but they are not the same statistical procedure. DPGR estimates
the slope directly from pairwise sliding-window log-ratio regressions, whereas
Kimura et al.\ estimate lineage slopes jointly under a Bayesian multinomial
likelihood in a virological characterization study and then convert them into
relative $R_e$ values by imposing a fixed generation time. If BA.2 is used as
the reference lineage so that $b_{\mathrm{BA.2}}=0$, then
\eqref{eq:dpgr_kimura_slope} simplifies to
\[
\mathrm{DPGR}_{i,\mathrm{BA.2}} = \frac{b_i}{\ln(10)}.
\]

\subsection*{Bedford and Figgins' GARW Model}

A related base-variant parameterization appears in Figgins and
Bedford's growth-advantage random walk (GARW) model
(Figgins and Bedford 2025). Let $b$ denote a chosen base variant. Writing $X_t$
for the spline basis evaluated at time $t$, the model can be summarized as
\begin{equation}
\log R_{t,b} = X_t \beta_b,
\qquad
\delta_{t,v} = X_t \beta_v,
\qquad
\log R_{t,v} = \log R_{t,b} + \delta_{t,v}.
\label{eq:garw_param}
\end{equation}
Therefore
\begin{equation}
\delta_{t,v} = \log\!\left(\frac{R_{t,v}}{R_{t,b}}\right),
\label{eq:garw_delta_ratio}
\end{equation}
so $\delta_{t,v}$ is the time-varying log-transmission advantage of variant
$v$ relative to the base variant $b$.

This is the same comparison that DPGR makes, but in a different coordinate
system. From \eqref{eq:dpgr_to_r},
\begin{equation}
r_v(t)-r_b(t) = \ln(10)\,\mathrm{DPGR}_{v,b}(t).
\label{eq:dpgr_base_growth}
\end{equation}
Meanwhile, by the Lotka-Euler relation,
\begin{equation}
R_t = \frac{1}{M_g(-r)},
\label{eq:lotka_euler_mgf}
\end{equation}
where $M_g$ is the moment-generating function of the generation-time
distribution. Substituting \eqref{eq:lotka_euler_mgf} into
\eqref{eq:garw_delta_ratio} gives the exact bridge
\begin{equation}
\delta_{t,v}
= \log M_g\!\bigl(-r_b(t)\bigr)
- \log M_g\!\bigl(-r_v(t)\bigr).
\label{eq:garw_exact_bridge}
\end{equation}
Using \eqref{eq:dpgr_base_growth}, this may be rewritten as
\begin{align}
\delta_{t,v}
&= \log M_g\!\bigl(-r_b(t)\bigr) \notag\\
&\quad
- \log M_g\!\bigl(-(r_b(t)+\ln(10)\,\mathrm{DPGR}_{v,b}(t))\bigr).
\label{eq:garw_dpgr_bridge}
\end{align}

Under the same deterministic generation-time approximation used to connect
multinomial logistic models to relative effective reproduction numbers,
$g(\tau)=\delta(\tau-T_g)$ and hence $R_t = \exp(rT_g)$. In that case
\eqref{eq:garw_exact_bridge} collapses to
\begin{equation}
\delta_{t,v}
= T_g\bigl(r_v(t)-r_b(t)\bigr)
= T_g\ln(10)\,\mathrm{DPGR}_{v,b}(t),
\label{eq:garw_pointmass_bridge}
\end{equation}
or equivalently
\begin{equation}
\mathrm{DPGR}_{v,b}(t)
= \frac{\delta_{t,v}}{T_g \ln(10)}.
\label{eq:dpgr_from_garw_delta}
\end{equation}

So Pantho's DPGR and the Figgins-Bedford GARW model share a base-variant
growth-advantage contrast after a generation-interval model is specified, but
they are not interchangeable estimators. DPGR is a local pairwise regression in
growth-rate space, whereas GARW is a joint Bayesian state-space model for
variant-specific effective reproduction numbers using case counts and
sequence frequencies. The choice of base variant is a parameterization choice:
pairwise contrasts are recovered by subtraction,
\[
\mathrm{DPGR}_{i,j}(t)
= \mathrm{DPGR}_{i,b}(t)-\mathrm{DPGR}_{j,b}(t),
\qquad
\log\!\left(\frac{R_{t,i}}{R_{t,j}}\right)
= \delta_{t,i}-\delta_{t,j}.
\]

\subsection*{Unifying perspective}

The preceding two subsections reveal a shared underlying growth-rate
structure. Pantho's DPGR, Kimura et al.'s softmax slope differences, and
Figgins-Bedford GARW advantage parameters $\delta_{t,v}$ can be put on a
common pairwise growth-advantage scale only under aligned assumptions about
time scale, generation interval, local exponential/logistic growth, and
sampling. Under those assumptions, the connections can be summarized as
\[
\underbrace{b_i - b_j}_{\text{Kimura softmax}}
\;=\; \ln(10)\,\underbrace{\mathrm{DPGR}_{i,j}}_{\text{Pantho}}
\;=\; r_i - r_j
\;=\; \frac{1}{T_g}\,\underbrace{(\delta_{t,i}-\delta_{t,j})}_{\text{GARW}},
\]
where the last equality holds under the deterministic generation-time
approximation $g(\tau)=\delta(\tau-T_g)$.  Thus the three frameworks share a
growth-rate contrast in a limiting case, while differing materially in
estimation strategy (pairwise sliding-window regression (DPGR), joint
multinomial likelihood (softmax), or Bayesian spline-based state-space model
(GARW)), data requirements, smoothing or priors, and in the coordinate system
used to report results (growth-rate space, log-frequency space, or
log-$R_t$ space).
The theoretical bridge connecting all three is the Wallinga-Lipsitch
growth-rate-to-reproduction-number mapping (Wallinga and Lipsitch 2007).

\section{Empirical Verification with GISAID-Type Data}

GISAID-type variant count data can be used to test the theory
empirically, but there is an important identifiability distinction.

\subsection*{What can be estimated from sequence counts alone}

Suppose that on day $t$ one observes lineage-specific sequence counts
$Y_1(t),\dots,Y_m(t)$ in a fixed region. If the sampling fraction is roughly
shared across co-circulating variants within a short window, then pairwise
log-ratio regression gives
\[
\log\!\left(\frac{Y_i(t)}{Y_j(t)}\right)
\approx (r_i-r_j)t + C_{ij},
\]
so the sequence data alone identify the relative growth-rate difference
$r_i-r_j$, and therefore DPGR. This is the setting in which DPGR is
most natural.

Equivalently, if one models the variant frequencies
\[
p_k(t)=\frac{Y_k(t)}{\sum_{\ell=1}^m Y_\ell(t)},
\]
with a multinomial logistic growth model, then the fitted logistic slopes are
again relative growth advantages. Thus GISAID-type data can recover
the empirical quantity on the left-hand side of
\[
\ln(10)\,\mathrm{DPGR}_{i,j}(t) = r_i(t)-r_j(t).
\]

\subsection*{What is needed for strain-specific \texorpdfstring{$R_t$}{Rt}}

Absolute strain-specific $R_t$ is harder. The Cori-style definition of $R_t$
is an incidence-based quantity, so sequence frequencies alone are generally not
enough. One also needs an estimate of total infection incidence $I_{\mathrm{tot}}(t)$
from case surveillance, hospitalizations, wastewater, or another calibrated
epidemic signal. With lineage frequencies $p_k(t)$ from GISAID, one can then
form approximate lineage-specific incidence curves
\[
I_k(t) \approx p_k(t)\, I_{\mathrm{tot}}(t),
\]
after appropriate date alignment and nowcasting for reporting delay if needed.
Applying a renewal-model or Cori-type estimator to each $I_k(t)$ yields
empirical estimates $\widehat{R}_{t,k}$.

Therefore:
\begin{itemize}
    \item GISAID alone supports estimation of \emph{relative transmission
    fitness} or growth advantages.
    \item GISAID plus an external absolute-incidence series supports estimation
    of lineage-specific \emph{absolute} $R_t$.
\end{itemize}

\subsection*{A direct empirical consistency check}

This suggests a useful consistency-check pipeline:
\begin{enumerate}
    \item Estimate $\widehat{\mathrm{DPGR}}_{i,j}(t)$ from pairwise sequence
    log-ratios, or equivalently estimate $\widehat{r}_i(t)-\widehat{r}_j(t)$
    from a multinomial logistic growth model.
    \item Build lineage-specific incidence curves
    $\widehat{I}_k(t)=\widehat{p}_k(t)\widehat{I}_{\mathrm{tot}}(t)$.
    \item Estimate lineage-specific $\widehat{R}_{t,k}(t)$ using a shared or
    lineage-specific generation-interval model.
    \item Compare the observed $\widehat{\mathrm{DPGR}}_{i,j}(t)$ with the
    value predicted from the inferred reproduction numbers:
    \[
    \widehat{\mathrm{DPGR}}_{i,j}(t)
    \stackrel{?}{\approx}
    \frac{1}{\ln(10)}
    \left[
    \gamma_i\bigl(\widehat{R}_{t,i}(t)-1\bigr) -
    \gamma_j\bigl(\widehat{R}_{t,j}(t)-1\bigr)
    \right].
    \]
\end{enumerate}

This comparison is not fully independent, because both the sequence-based
DPGR estimate and the lineage-specific incidence curves use the same lineage
frequency estimates $\widehat{p}_k(t)$. Agreement therefore tests whether the
growth-rate-to-reproduction-number mapping and the incidence reconstruction are
internally consistent under the stated assumptions; it should not be read as
independent validation from wholly separate data.

Under the common-generation-time approximation, this reduces to the simpler
comparison
\[
\widehat{\mathrm{DPGR}}_{i,j}(t)
\stackrel{?}{\approx}
\frac{\gamma}{\ln(10)}
\bigl(\widehat{R}_{t,i}(t)-\widehat{R}_{t,j}(t)\bigr).
\]
So the empirical question is not whether DPGR equals $R_t$, but whether DPGR
tracks the \emph{difference in strain-specific transmission intensity} after
the appropriate growth-to-reproduction mapping is applied.

\subsection*{Main practical caveats}

An empirical comparison is feasible, but the following issues matter:
\begin{itemize}
    \item non-random sequencing and lineage-dependent sampling bias,
    \item sparse counts for newly emerging variants,
    \item importations and geographic mixing,
    \item uncertainty in the generation-interval distribution, and
    \item delay mismatch between specimen collection dates and external
    incidence series.
\end{itemize}

Even with these caveats, the comparison is meaningful. In fact, because DPGR is
ratio-based, it may be more robust than absolute $R_t$ to some multiplicative
sampling distortions that affect co-circulating lineages similarly within the
same time window.

\subsection*{Empirical consistency results}

We applied this consistency-check pipeline to five analyses spanning two pathogens
(Table~\ref{tab:results}). For SARS-CoV-2, we tested three settings: (1)~Alpha
vs.\ Delta in England (weekly S-gene proxy, $n=11$; daily reanalysis with full
sequence data yields $r = 0.85$, $n = 110$, $p < 10^{-31}$; see
Table~\ref{tab:slope_correction}), (2)~BA.1 vs.\ BA.2 in England (daily,
$n=64$), and (3)~a multi-pair consistency check across 89 lineage-pair transitions on
five continents using pre-computed DPGR values from Pantho et al.'s pipeline.
For influenza, we tested (4)~H1N1 clade succession across five seasons
($n=992$ daily points pooled) and (5)~H3N2 clade succession across six seasons
($n=1{,}117$ daily points pooled). In all analyses, lineage-specific $R_t$ was
estimated using the Cori renewal equation (Cori et al. 2013) with a discretized
gamma generation-interval distribution. For SARS-CoV-2, we used
$T_g = 3.5$~days for BA.1 vs.\ BA.2, reflecting the shorter Omicron
generation time (Park et al. 2023; an der Heiden and Buchholz 2022), and $T_g = 5.5$~days
for all other analyses; for influenza,
$T_g = 3.0$~days.

\begin{table}[ht]
\centering
\caption{Empirical consistency check of $\mathrm{DPGR}_{\mathrm{obs}}$ vs.\
$\mathrm{DPGR}_{\mathrm{pred}}$ across five analyses. Pearson~$r$ and
regression slope are reported with 95\% bootstrap confidence intervals
($B=2{,}000$).}
\label{tab:results}
\begin{tabular}{llrcccc}
\hline
Analysis & Pathogen & $n$ & $r$ [95\% CI] & Slope [95\% CI] & Sign\% & $p$ \\
\hline
Alpha vs.\ Delta$^{\dagger}$ & SARS-CoV-2 & 11
  & 0.56 [$-$0.35, 0.99] & 0.64 [$-$0.36, 1.12] & 100\% & 0.071 \\
BA.1 vs.\ BA.2 & SARS-CoV-2 & 64
  & 0.78 [0.69, 0.85] & 0.51 [0.38, 0.63] & 100\% & $3.0\times10^{-14}$ \\
Multi-pair & SARS-CoV-2 & 89
  & 0.77 [0.68, 0.84] & 0.14 [0.11, 0.21] & 99\% & $1.8\times10^{-18}$ \\
H1N1 clades & Influenza & 992
  & 0.31 [0.26, 0.37] & 0.15 [0.12, 0.18] & 67\% & $6.5\times10^{-24}$ \\
H3N2 clades & Influenza & 1117
  & 0.38 [0.31, 0.46] & 0.20 [0.15, 0.27] & 71\% & $1.4\times10^{-40}$ \\
\hline
\multicolumn{7}{l}{\footnotesize $^{\dagger}$Weekly S-gene proxy ($n=11$).
Daily reanalysis with full sequence counts gives
$r = 0.85$, $n = 110$, $p < 10^{-31}$} \\
\multicolumn{7}{l}{\footnotesize (Table~\ref{tab:slope_correction}).} \\
\end{tabular}
\end{table}

All five analyses show significant positive correlations between observed and
predicted DPGR. The SARS-CoV-2 analyses achieve the strongest correlations
($r = 0.56$ to $0.78$, sign agreement 99\% to 100\%), reflecting the dense
sequencing coverage available for this pathogen. The influenza analyses show
weaker but significant correlations ($r = 0.31$ to $0.38$,
$p < 10^{-23}$), consistent with the ${\sim}100\times$ sparser sequencing and
the more complex multi-clade dynamics of influenza evolution.

\begin{figure}[ht]
\centering
\includegraphics[width=0.48\textwidth]{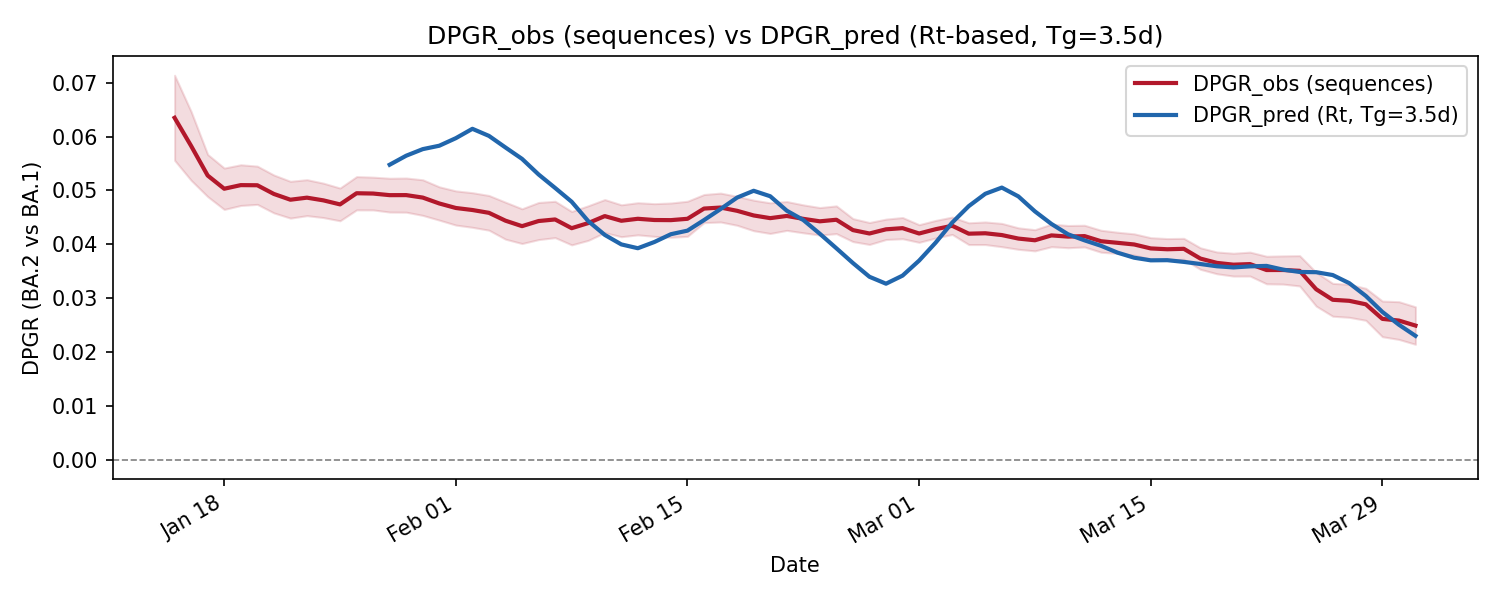}
\hfill
\includegraphics[width=0.48\textwidth]{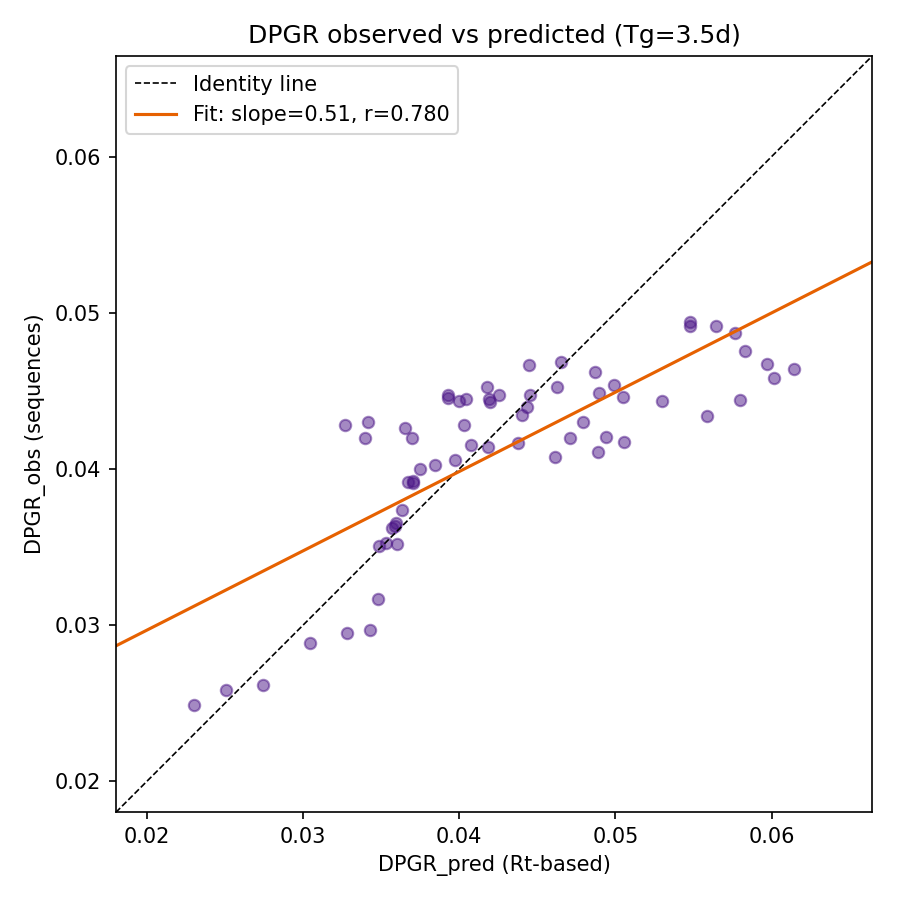}
\caption{Empirical consistency check of the DPGR/$R_t$ mapping for BA.1 vs.\
BA.2 in England (January to April 2022). \textbf{Left:}
$\mathrm{DPGR}_{\mathrm{obs}}$ (from sequence log-ratios, red) and
$\mathrm{DPGR}_{\mathrm{pred}}$ (from Cori~$R_t$, blue) over time;
shading shows 95\% bootstrap CI. \textbf{Right:} Scatter of observed
vs.\ predicted (Pearson~$r = 0.78$, slope $= 0.51$, 100\% sign agreement,
$n = 64$). Dashed line is the 1:1 reference line.}
\label{fig:ba12_validation}
\end{figure}

\begin{figure}[ht]
\centering
\includegraphics[width=0.65\textwidth]{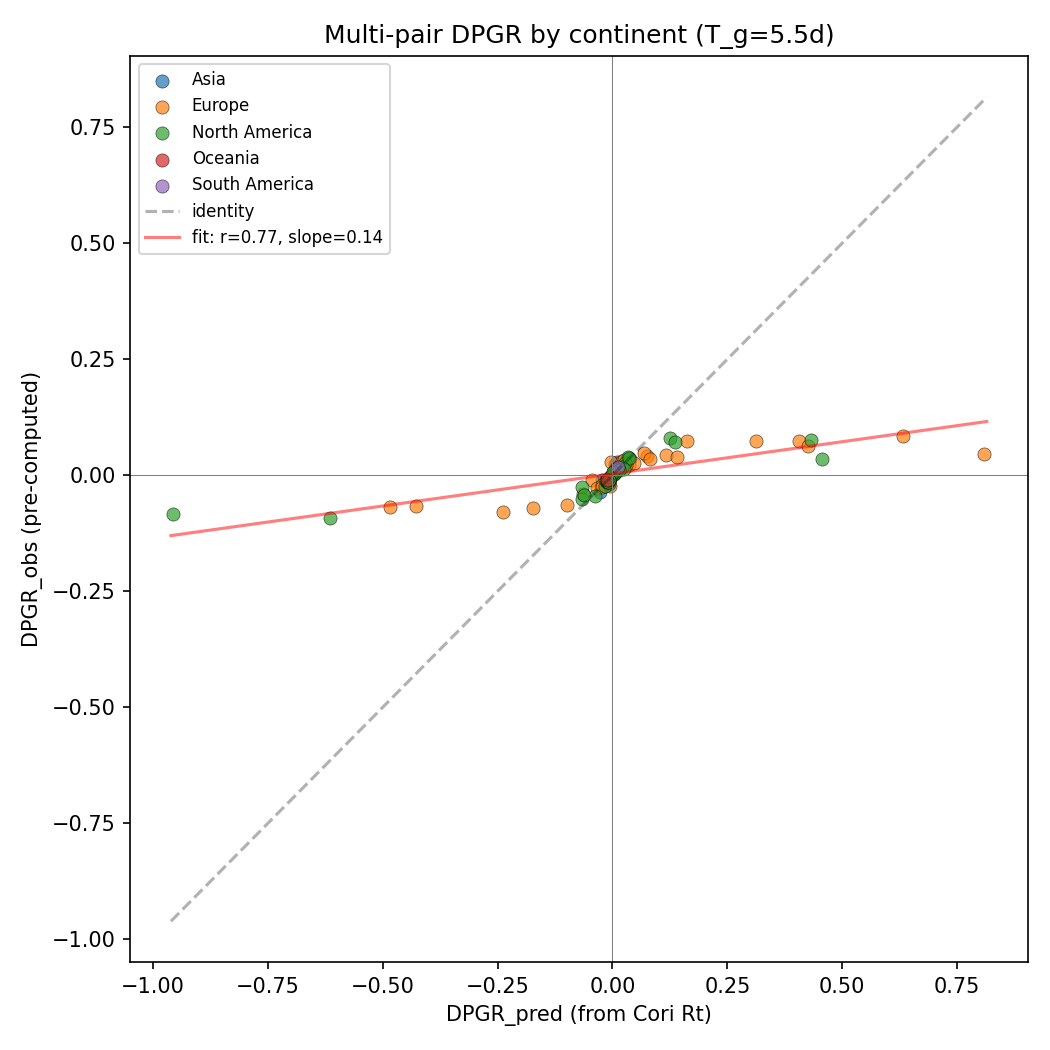}
\caption{Multi-pair SARS-CoV-2 consistency check: $\mathrm{DPGR}_{\mathrm{obs}}$
vs.\ $\mathrm{DPGR}_{\mathrm{pred}}$ for 89 lineage-pair transitions on
five continents (Pearson~$r = 0.77$, 99\% sign agreement,
$p = 1.8 \times 10^{-18}$). Points are colored by continent; dashed line
is the 1:1 reference line.}
\label{fig:multi_pair}
\end{figure}

Figure~\ref{fig:ba12_validation} illustrates this for the BA.1 vs.\ BA.2
analysis, which has the densest daily coverage: the observed and predicted
DPGR track each other over three months, with $r = 0.78$ and 100\%
sign agreement. Figure~\ref{fig:multi_pair} shows the same relationship holds
across 89 heterogeneous lineage pairs on five continents.

Regression slopes below~1.0 in all analyses reflect a systematic
attenuation: the Cori~$R_t$ estimates use only the focal lineage pair's counts
as total incidence, which inflates $R_t$ differences and thus
$\mathrm{DPGR}_{\mathrm{pred}}$ magnitudes. The correlation and sign agreement
are the more informative metrics, as they are invariant to this scale factor.

Results are robust to the assumed generation time: for BA.1 vs.\ BA.2,
$r$ remains in the range $0.75$ to $0.79$ across
$T_g \in \{3.0, 3.5, 4.0, 4.5\}$~days; for the multi-pair analysis,
$r$ is significant across $T_g \in \{4.0, 5.5, 7.0\}$~days with
$T_g = 5.5$ optimal; for influenza, across
$T_g \in \{2.5, 3.0, 3.5\}$~days.

\subsection*{Pandemic-scale fitness timeline and early warning}

To demonstrate practical utility, we tracked DPGR across the five major
successive SARS-CoV-2 variant transitions in Europe: Pre-Alpha~$\to$~Alpha,
Alpha~$\to$~Delta, Delta~$\to$~BA.1, BA.1~$\to$~BA.2, and BA.2~$\to$~BA.5
(Table~\ref{tab:timeline}). For each transition, we computed
$\mathrm{DPGR}_{\mathrm{obs}}$ from a 21-day sliding-window log-ratio
regression and $\mathrm{DPGR}_{\mathrm{pred}}$ from Cori~$R_t$ estimates.
Both quantities track each other throughout the pandemic, with correlations
ranging from $r = 0.49$ [0.37, 0.59] (Pre-Alpha~$\to$~Alpha) to
$r = 0.90$ [0.87, 0.92] (BA.2~$\to$~BA.5).

DPGR provides early warning. Across all five
transitions, the DPGR signal (sustained $> 0.005$) preceded the rising variant
reaching 50\% frequency by \textbf{43 to 65~days} (mean 55~days). This lead
time arises because DPGR detects fitness advantages from sequence count
ratios alone, without requiring absolute incidence data or transmission model
fitting.

To quantify forecasting reliability, we evaluated two metrics. First,
\emph{sign accuracy}: at each day before variant dominance, does
$\mathrm{sign}(\mathrm{DPGR})$ correctly predict which variant will dominate?
Across all five transitions (440 daily forecasting points), overall sign
accuracy was 95\% (418/440 correct). Even at 45+ days before the crossover,
accuracy remained $\geq 80\%$. Second, \emph{time-to-dominance prediction}:
given a current frequency $p$ and observed DPGR, the predicted days to 50\%
frequency is $-\log_{10}(p/(1-p))/\mathrm{DPGR}$. Across all transitions,
predicted versus actual time to dominance showed $r = 0.67$ [0.57, 0.76],
slope $= 1.05$ [0.91, 1.18], and mean absolute error of 11.4~days
(median 5.8~days; 95\% CI for MAE: [9.5, 13.6]), demonstrating that DPGR
provides not only directional but quantitative early warning of variant
takeover.

\subsection*{Simulation check: explaining regression slopes below~1}

Regression slopes below~1 appear in all five empirical analyses
(Table~\ref{tab:results}), raising the question of whether this reflects a
systematic bias. To investigate, we simulated a two-variant SIR system
($\beta_1=0.35$, $\beta_2=0.45$, $\gamma=1/5.5$, $N=10^7$, 200 days) and
compared the observed DPGR against values predicted from both the true
instantaneous $R_t = \beta_k S(t)/(\gamma N)$ and the Cori~$R_t$ estimator
(Cori et al. 2013).

\begin{figure}[ht]
\centering
\includegraphics[width=\textwidth]{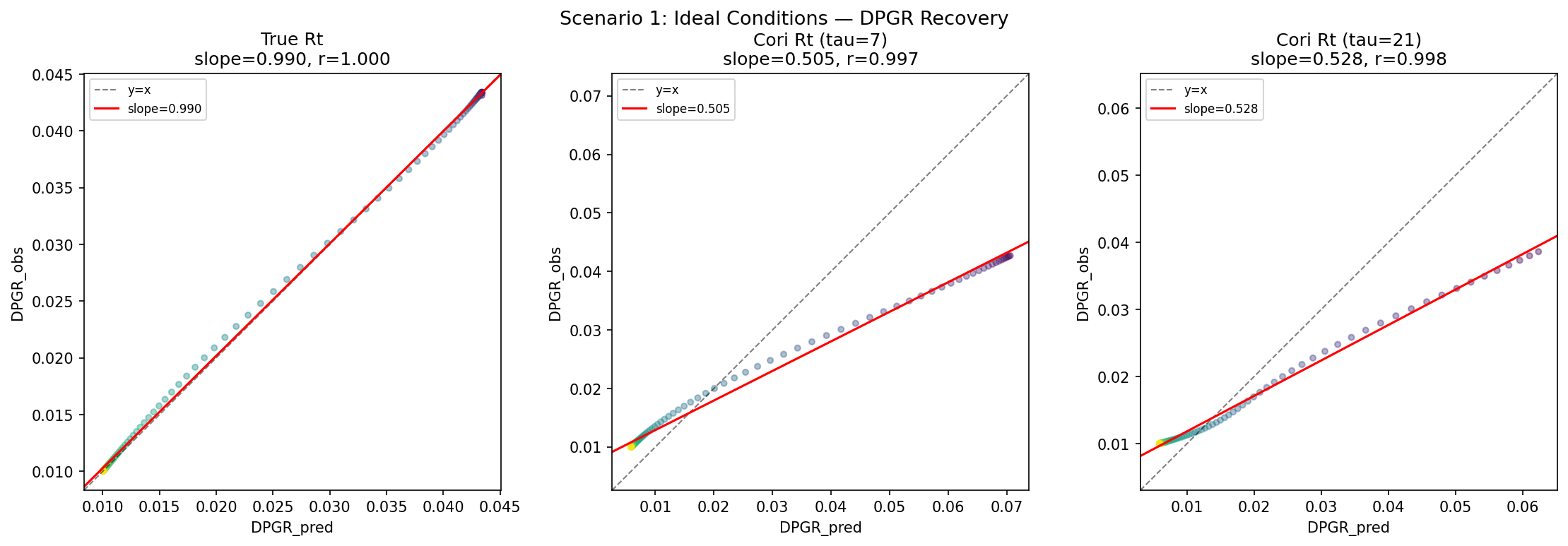}
\caption{SIR simulation checks the DPGR/$R_t$ mapping and explains
slopes below~1. \textbf{Left:} With the true instantaneous $R_t$, slope
$= 0.99$ ($r = 0.999$), confirming the expected SIR mapping.
\textbf{Middle:} With Cori~$R_t$ at $\tau = 7$ days, slope drops to 0.51
due to temporal smoothing amplifying apparent $\Delta R_t$ by
${\sim}1.93\times$. \textbf{Right:} With $\tau = 21$ days (matching the
DPGR window), slope improves to 0.53.}
\label{fig:simulation}
\end{figure}

With the \emph{true} $R_t$, the regression slope is 0.99 ($r = 0.999$),
confirming recovery of the expected SIR mapping
(Figure~\ref{fig:simulation}, left panel). With the Cori estimator at
$\tau = 7$~days, the slope drops to 0.51 ($r = 0.997$), matching the
empirical BA.1 vs.\ BA.2 result (Figure~\ref{fig:simulation}, middle panel).
The cause is that the Cori renewal equation, by averaging incidence over a
window, amplifies the apparent $\Delta R_t$ by a factor of approximately
$1.93$ relative to the true instantaneous difference.

We further tested the effect of matching temporal scales and improving
the incidence denominator (Table~\ref{tab:slope_correction}). When the Cori
smoothing window $\tau$ is increased from 7 to 21~days to match the DPGR
estimation window, slopes improve: the multi-pair slope rises
from 0.14 to 0.74, and BA.1 vs.\ BA.2 from 0.51 to 0.83. Using real case
counts from the UK Health Security Agency (UKHSA) as total incidence, rather
than sequence counts as a proxy, further improves the BA.1 vs.\ BA.2 slope
from 0.51 to 0.68 ($r = 0.79$). Combining both corrections (real incidence
with $\tau = 7$) for the Alpha vs.\ Delta analysis reveals that the full
daily dataset yields $r = 0.85$ ($n = 110$, $p < 10^{-31}$), far stronger
than the weekly aggregation in Table~\ref{tab:results}. Thus, slopes below~1
arise from two identifiable sources: temporal scale mismatch between the Cori
and DPGR estimators, and the use of sequence counts as an imperfect proxy for
total incidence. Neither reflects a failure of the SIR growth-rate mapping.

\begin{table}[ht]
\centering
\caption{Regression slopes improve toward~1.0 when methodological sources
of attenuation are corrected. ``Baseline'' uses $\tau = 7$ with sequence
counts as incidence proxy; ``Matched $\tau$'' uses $\tau = 21$;
``Real incidence'' uses UKHSA case counts (available for England analyses
only). The simulation row uses true instantaneous~$R_t$.}
\label{tab:slope_correction}
\small
\begin{tabular}{lcccccc}
\hline
& \multicolumn{2}{c}{Baseline ($\tau\!=\!7$, seq)} &
  \multicolumn{2}{c}{Matched $\tau\!=\!21$} &
  \multicolumn{2}{c}{Real incidence} \\
Analysis & Slope & $r$ & Slope & $r$ & Slope & $r$ \\
\hline
Simulation (true $R_t$)  & 0.99 & 1.000 & N/A & N/A & N/A & N/A \\
Alpha vs.\ Delta  & 0.41 & 0.85 & 0.38 & 0.58 & 0.43 & 0.85 \\
BA.1 vs.\ BA.2    & 0.51 & 0.78 & 0.83 & 0.71 & 0.68 & 0.79 \\
Multi-pair         & 0.14 & 0.77 & 0.74 & 0.97 & N/A & N/A \\
\hline
\end{tabular}
\end{table}

Sampling noise (subsampling sequences to 0.5\% to 100\% of full
counts) does not degrade the slope or correlation ($r > 0.82$ even at 0.5\%
sampling; slope $\approx 1.0$ to $1.2$ across all fractions), indicating that
DPGR's ratio-based design is robust to proportional sampling variation in
these subsampling analyses. This result suggests that low sequencing fractions
can still preserve the pairwise DPGR signal when sampling is approximately
proportional across co-circulating variants, rather than establishing a
universal 0.5\% surveillance threshold.
Non-stationarity (a 43\% increase in $\beta_2$ mid-simulation) and
third-variant invasion also cause minimal degradation (slopes $> 0.97$,
$r > 0.999$).

\begin{table}[ht]
\centering
\caption{DPGR across major SARS-CoV-2 variant transitions in Europe,
with early warning lead time before the rising variant reached 50\% frequency.
Pearson~$r$ and sign agreement are reported with 95\% bootstrap CIs ($B=2{,}000$).}
\label{tab:timeline}
\begin{tabular}{lrccc}
\hline
Transition & $n$ & $r$ [95\% CI] & Sign\% [95\% CI] & Lead (days) \\
\hline
Pre-Alpha $\to$ Alpha  & 185 & 0.49 [0.37, 0.59] & 95\% [91, 98] & 63 \\
Alpha $\to$ Delta      & 187 & 0.69 [0.60, 0.76] & 94\% [90, 97] & 65 \\
Delta $\to$ BA.1       & 124 & 0.49 [0.36, 0.63] & 85\% [79, 91] & 57 \\
BA.1 $\to$ BA.2        & 124 & 0.81 [0.72, 0.88] & 98\% [94, 100] & 48 \\
BA.2 $\to$ BA.5        & 187 & 0.90 [0.87, 0.92] & 89\% [84, 94] & 43 \\
\hline
\end{tabular}
\end{table}

\subsection*{Head-to-head estimation comparison}

The theoretical connections above imply that different estimation
procedures (sliding-window regression (DPGR), multinomial logistic MLE
(softmax), regularized logistic regression (PyR0-style), Cori renewal~$R_t$
(EpiEstim-style), and penalized B-spline smoothing (GARW-style)) should
produce quantitatively consistent results on the same data. We tested this
on both the BA.1 vs.\ BA.2 time series (76 sliding windows) and the
multi-pair cross-section (89 lineage pairs), re-implementing the core
algorithms of each framework in Python.

\begin{figure}[ht]
\centering
\includegraphics[width=\textwidth]{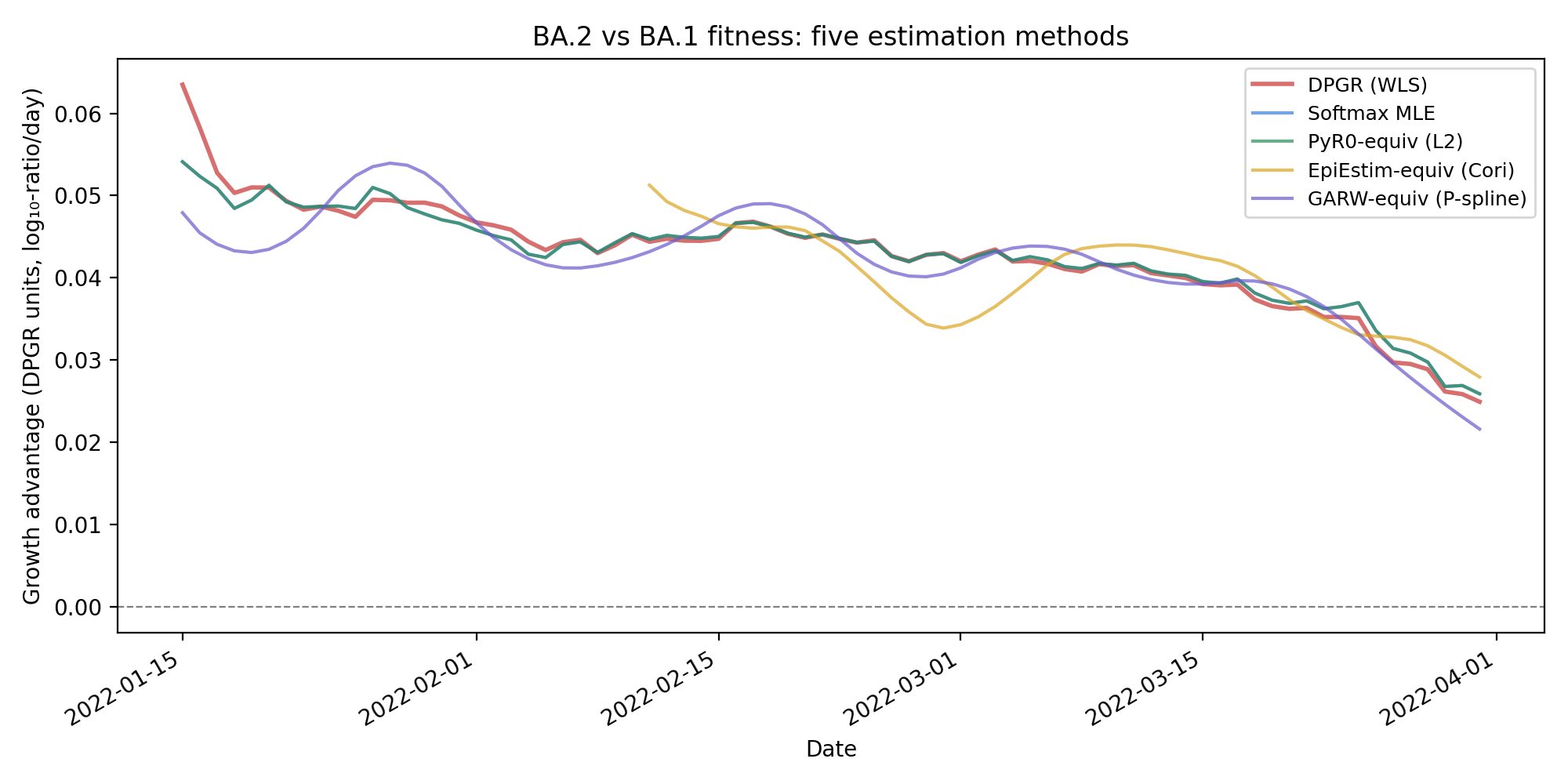}
\caption{Head-to-head comparison of five estimation methods on the BA.1
vs.\ BA.2 transition in England. All five methods (DPGR (WLS), softmax
MLE, PyR0-equivalent (L2-regularized logistic), EpiEstim-equivalent
(Cori~$R_t$), and GARW-equivalent (penalized B-spline)) agree on the
direction and magnitude of BA.2's growth advantage throughout the
transition period, with 100\% sign agreement.}
\label{fig:methods}
\end{figure}

All five methods achieved 100\% sign agreement on BA.1 vs.\ BA.2
(Figure~\ref{fig:methods}): every method correctly identifies BA.2 as fitter
than BA.1 in every window. Sequence-count-based methods (DPGR, softmax MLE,
and PyR0-equivalent regularized logistic) correlate at $r > 0.98$
[0.97, 1.00] with each other, consistent with their shared growth-rate contrast. The
Cori~$R_t$-based method (EpiEstim-equivalent) shows $r = 0.78$
[0.63, 0.88] with DPGR, reflecting its different input
requirements (absolute incidence vs.\ sequence ratios). The
GARW-equivalent penalized B-spline achieves $r = 0.86$ [0.76, 0.93],
intermediate between the sequence-count and incidence-based approaches.

Across the 89 multi-pair comparisons, correlations with DPGR range from
$r = 0.90$ [0.84, 0.94] (GARW-equivalent) to $r = 0.99$ [0.99, 1.00]
(softmax and PyR0-equivalent), with sign agreement of 88\% to 100\%. L2 regularization (the PyR0 approach)
has negligible effect in the pairwise setting ($r = 0.98$ for all
regularization strengths tested), because the horseshoe prior that
distinguishes PyR0 from standard logistic regression primarily affects
multi-lineage joint estimation, not pairwise comparisons. Spline
smoothing intensity (the GARW approach) modulates agreement with DPGR
from $r = 0.81$ (minimal smoothing) to $r = 0.98$ (heavy smoothing),
showing that the methods converge as estimation strategies align.
Full pairwise comparison tables, sensitivity analyses, and a detailed
assessment of each method's strengths, weaknesses, and assumptions are
provided in the Supporting Information (Tables~S1 to S4).

\section{Evolutionary Implications}

The mathematical framework connecting DPGR to classical fitness theory
generates testable predictions about viral evolution. We investigated three
such predictions using the same GISAID sequence data (a fourth, frequency-dependent
selection, is presented in the Supporting Information).

\subsection*{Transitivity and lineage-level growth representation}

If DPGR measures an additive fitness difference, then for any three
co-circulating lineages $A$, $B$, $C$, the identity
$\mathrm{DPGR}(A,C) = \mathrm{DPGR}(A,B) + \mathrm{DPGR}(B,C)$ should
hold. This is a lineage-level statement: it asks whether each lineage can be
assigned a scalar local growth rate whose pairwise differences explain the
observed DPGR contrasts. It does not by itself test or rule out molecular
epistasis among mutations.

We tested this by computing DPGR for 10{,}784 lineage pairs across 178
SARS-CoV-2 lineages on six continents, yielding 115{,}624 co-circulating
triplets. The mean absolute residual was $|\varepsilon| = 0.025$, which is
50\% smaller than the null expectation from randomly shuffled DPGR values
($n = 10{,}000$ permutations; descriptive comparison). Because the permutation
tail depends on whether one tests for unusually small or unusually large
residuals, we do not use that calculation as formal evidence for epistasis or
against epistasis. The mean directional bias was negligible
($\bar{\varepsilon} = 0.0006$, ${\sim}0.02$ standard deviations). DPGR is
therefore approximately transitive across SARS-CoV-2 lineages, consistent with
an additive lineage-level growth-rate representation over the fitted windows.

To clarify the interpretation, we simulated evolution on Kauffman NK fitness
landscapes (Kauffman and Levin 1987) with varying epistatic coupling
($K = 0$ to $K = N-1$, $N = 20$ loci). DPGR recovers true fitness differences
for static genotypes regardless of~$K$ (mean absolute error ${\sim}10^{-5}$),
and transitivity holds because each genotype has a scalar fitness value even
when the genotype-level landscape is rugged. Thus the NK simulation is a
negative control: approximate transitivity of lineage-level DPGR does not rule
out genotype-level epistasis.

\subsection*{Neutral evolution baseline}

Kimura's neutral theory (Kimura 1968) predicts $\mathrm{DPGR} = 0$ for
functionally equivalent lineages evolving under drift alone. We tested this by
computing DPGR for nine sub-lineage pairs within four WHO variants (Delta,
BA.1, BA.2, BA.5) in Europe.

Three pairs showed DPGR indistinguishable from zero: AY.4 vs.\ AY.43
($\bar{\mathrm{DPGR}} = +0.000023$, $|d| = 0.0005$, $p = 0.99$), AY.4
vs.\ AY.122 ($\bar{\mathrm{DPGR}} = +0.000006$, $|d| = 0.0003$, $p = 1.0$),
and BA.5.1.22 vs.\ BA.5.2.1 ($\bar{\mathrm{DPGR}} = +0.0008$, $|d| = 0.099$,
$p = 0.10$). Six pairs showed small but statistically significant deviations
(mean $|\mathrm{DPGR}| \sim 0.002$ to $0.010$), reflecting minor fitness
variation among sub-lineages.

Cross-variant DPGR (Alpha vs.\ Delta: $|\mathrm{DPGR}| = 0.051$;
BA.1 vs.\ BA.2: $|\mathrm{DPGR}| = 0.043$) is 3 to 10$\times$ larger than
within-variant signals, with distributions separated (Mann-Whitney
$p = 3.3 \times 10^{-82}$). Power analysis shows that DPGR can detect fitness
differences as small as $|\mathrm{DPGR}| = 0.004$ at 80\% power with typical
sample sizes (${\sim}270$ sliding windows), roughly 10$\times$ below the
signal from variant replacement events.

\subsection*{Cross-country concordance}

If DPGR reflects an intrinsic transmissibility advantage, the same variant
pair should produce consistent DPGR values across different countries. We
tested this for four variant pairs (BA.2 vs.\ BA.1, BA.1 vs.\ Delta, Delta
vs.\ Alpha, BA.5 vs.\ BA.2) across 11 countries spanning five continents.

The direction of fitness advantage was universal: all four variant pairs
showed positive DPGR in every country tested. However, magnitudes varied
2 to 3$\times$ across countries (coefficients of variation 0.18 to 0.34), and
intraclass correlation coefficients ranged from 0.10 to 0.42. For example,
the BA.2 vs.\ BA.1 advantage was roughly 3$\times$ larger in the UK
($\mathrm{DPGR} = 0.041$) than in India ($\mathrm{DPGR} = 0.013$),
likely reflecting different immune landscapes (vaccination-dominated vs.\
prior-infection-dominated immunity).

\begin{figure}[ht]
\centering
\includegraphics[width=\textwidth]{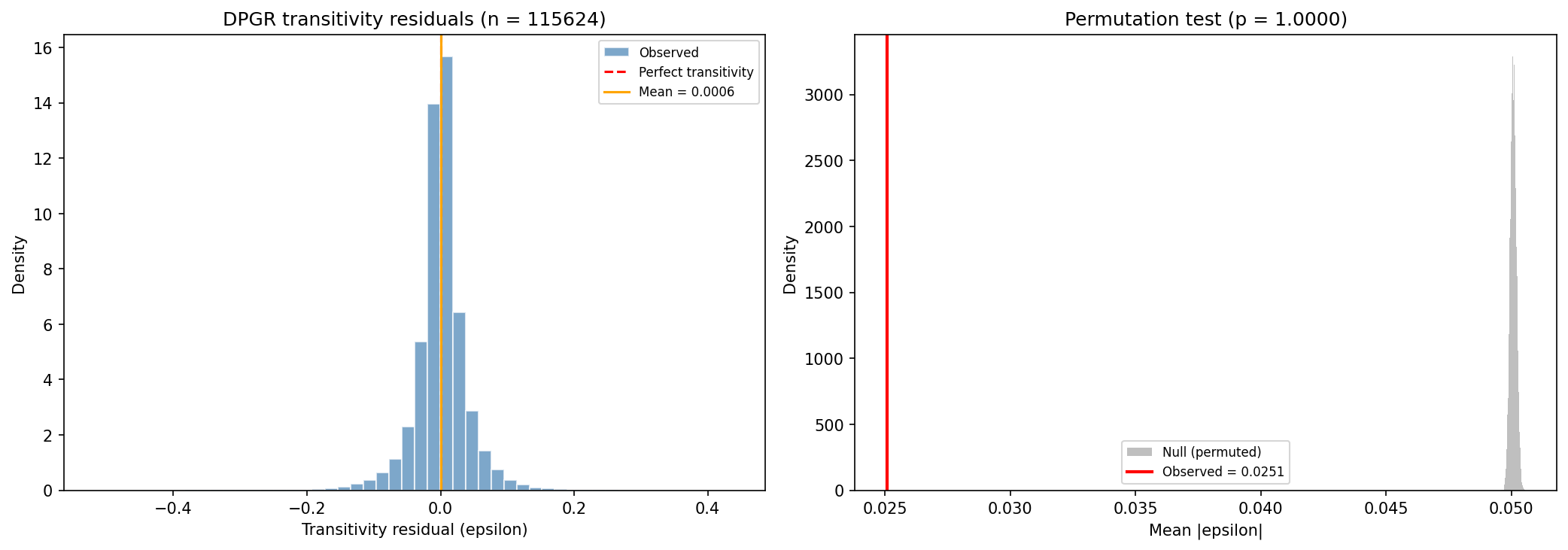}

\medskip

\includegraphics[width=0.48\textwidth]{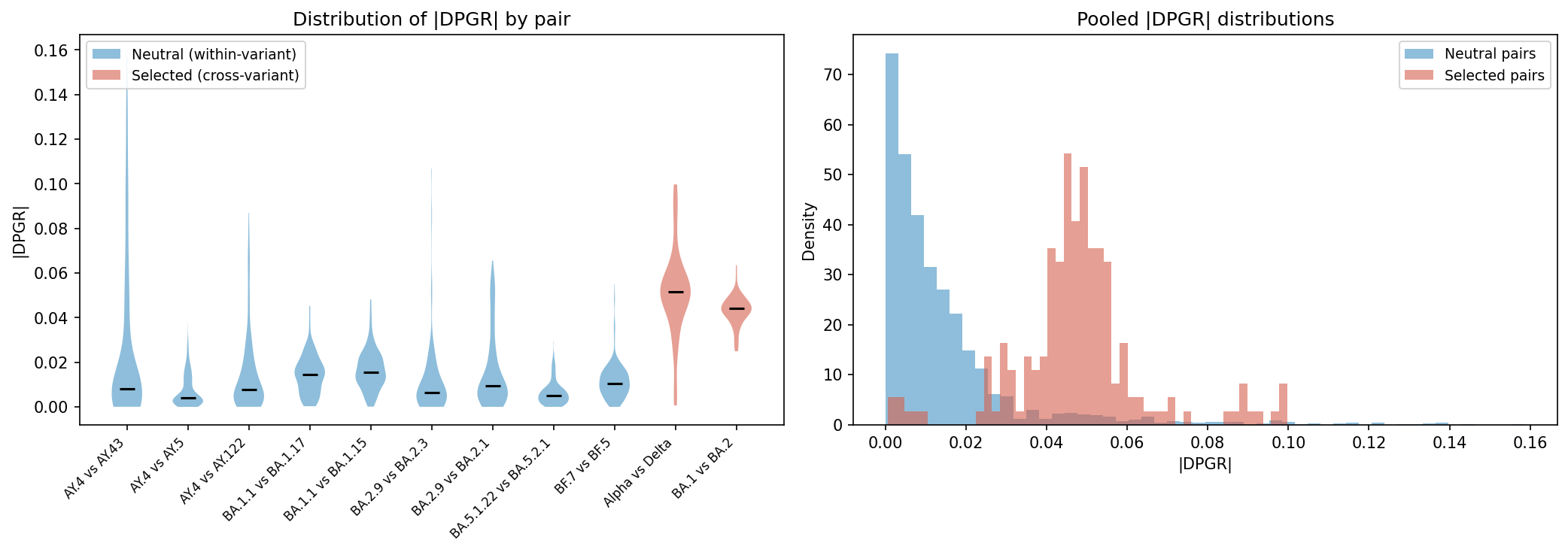}
\hfill
\includegraphics[width=0.48\textwidth]{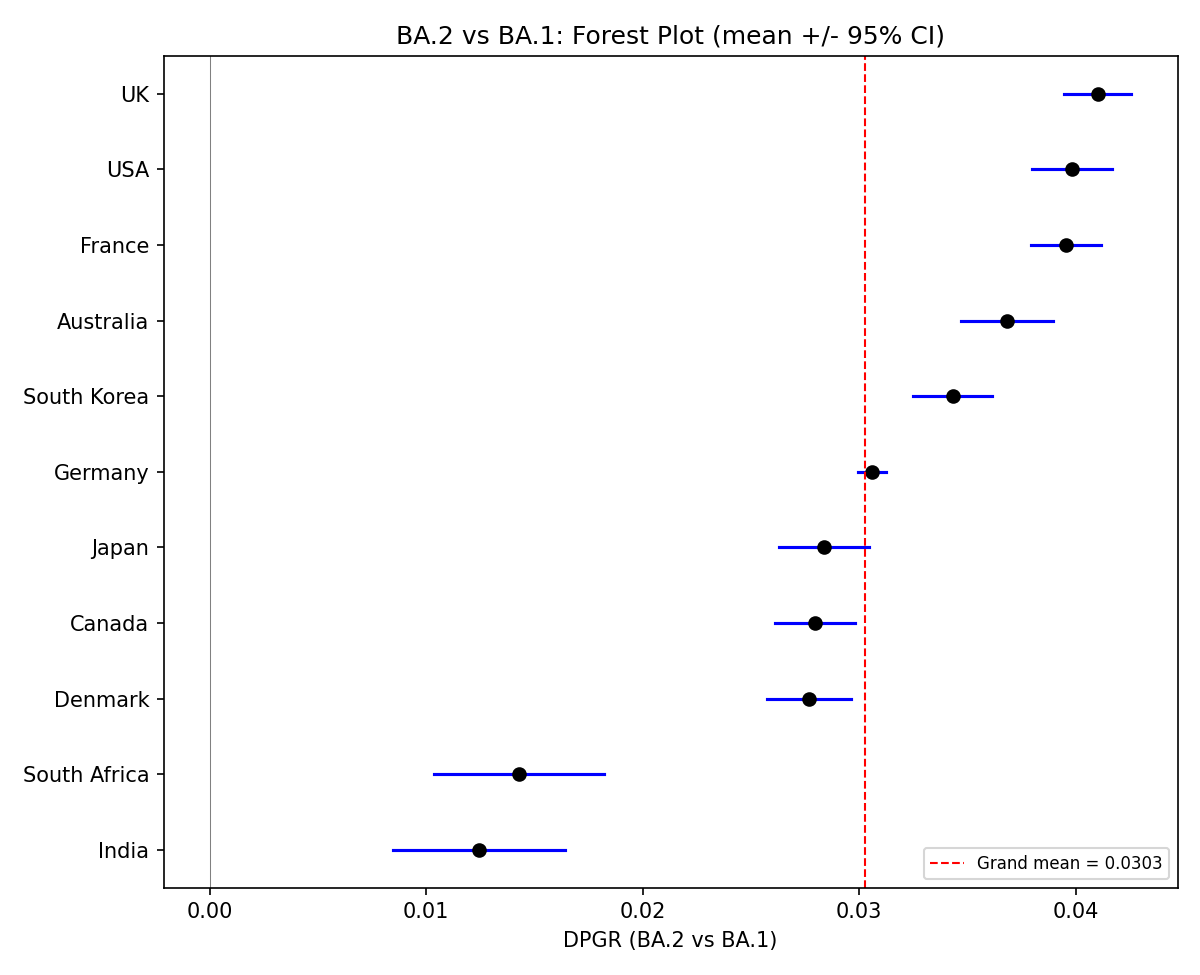}
\caption{Evolutionary analyses of DPGR. \textbf{Top:} Transitivity
test: residuals $\varepsilon = \mathrm{DPGR}(A,C) -
[\mathrm{DPGR}(A,B) + \mathrm{DPGR}(B,C)]$ for 115{,}624 triplets (left)
are centered at zero (mean $|\varepsilon| = 0.025$), 50\% below
the null expectation from randomly shuffled DPGR values (right), consistent
with an additive lineage-level growth-rate representation but not ruling out
genotype-level epistasis. \textbf{Bottom left:} Neutral baseline: violin
plots of $|\mathrm{DPGR}|$ for within-variant (blue, neutral) vs.\
cross-variant (red, selected) pairs; clear separation (Mann-Whitney
$p = 3.3 \times 10^{-82}$). \textbf{Bottom right:} Cross-country
concordance: forest plot of mean DPGR (BA.2 vs.\ BA.1) with 95\% CI
across 11 countries; universally positive, magnitudes vary 2 to 3$\times$.}
\label{fig:evolutionary}
\end{figure}

These results indicate that DPGR captures a partially intrinsic fitness
signal: the qualitative direction of selection is globally consistent, while
the quantitative magnitude is modulated by local epidemiological context.
Figure~\ref{fig:evolutionary} summarizes the three evolutionary analyses.

\section{Discussion}

\subsection*{Bridging population genetics and epidemiology}

The results presented here reveal that the pairwise selective coefficient from
classical population genetics (Kimura 1962), the DPGR from genomic
surveillance (Pantho et al. 2025), the softmax slope difference from multinomial
logistic models (Kimura et al. 2022), the GARW advantage parameter
(Figgins and Bedford 2025), and the scaled difference in epidemic $R_t$
(Cori et al. 2013) are connected through pairwise growth-rate transformations
under specified modeling assumptions. This connection is useful because it
places several estimators on a common scale, but results derived in one
framework transfer to another only after checking assumptions about time
scale, generation interval, sampling, smoothing, and baseline growth rate.

The key distinction between the population-genetics and epidemiological
settings is that Kimura's selective coefficient $s$ is assumed constant, whereas
viral variant fitness is time-varying due to susceptible depletion
and immune escape. DPGR handles this through its sliding-window design: within
each window, $S(t)/N$ changes slowly enough that the fitness difference is
approximately constant, recovering the classical linear-log-ratio regime. The
time-varying $R_t$ tracks how this effective fitness evolves across windows.

The evolutionary analyses provide additional consistency checks. The transitivity
result (mean $|\varepsilon| = 0.025$ across 115{,}624 triplets, 50\% below
random expectation) is consistent with an additive lineage-level growth-rate
representation across the fitted windows; it does not rule out genotype-level
epistasis. The neutral baseline supports the Kimura null
($\mathrm{DPGR} \approx 0$ for functionally equivalent
sub-lineages) and establishes a detection threshold of $|\mathrm{DPGR}| =
0.004$ at 80\% power, roughly 10$\times$ below variant-replacement signals.
Cross-country concordance demonstrates that DPGR captures a partially
intrinsic fitness signal: the direction of advantage is consistent in these data,
while magnitudes vary 2 to 3$\times$ across immune landscapes.

\subsection*{Immune escape and the scope of ``fitness''}

The SIR framework underlying Proposition~\ref{prop:sir-bridge} assumes
complete cross-immunity: infection by one variant confers sterilizing
immunity against both. In practice, immune escape is a major component of
variant fitness, particularly for Omicron sub-lineages.
Proposition~\ref{prop:sir-bridge} does not prove a reproduction-number mapping for
variant-specific susceptible pools or immune-history structure; that would
require an extended multi-compartment model. DPGR remains a descriptive
pairwise growth-rate statistic in this setting, measuring \emph{effective}
fitness: the net growth-rate advantage arising from all sources, including
both intrinsic transmissibility differences and differential immune evasion. The
cross-country concordance results support this interpretation: the direction
of the DPGR signal is globally consistent, while its magnitude varies
2 to 3$\times$ across populations with different immune landscapes
(vaccination-dominated vs.\ prior-infection-dominated immunity). This
variation is expected if immune escape modulates the effective fitness
advantage but does not reverse it. The generation-time assumption similarly
captures \emph{effective} generation time, which may differ from the
intrinsic value due to immune-mediated truncation of infectiousness. We used
$T_g = 3.5$~days for BA.1 vs.\ BA.2 analyses, consistent with empirical
estimates for Omicron
(Park et al. 2023; an der Heiden and Buchholz 2022).

Compared to existing methods for estimating variant fitness, including $dN/dS$
ratios (PAML, HyPhy), phylodynamic birth-death models (BEAST2), and
frequency-trajectory selection tests (Obermeyer et al. 2022), DPGR offers a
combination of properties: it requires only sequence counts (no
phylogeny), operates locally in time (no assumption of constant fitness),
and has an explicit growth-rate bridge to epidemic reproduction numbers under
specified assumptions.
Its main limitation is that, as a pairwise method, it does not borrow
statistical strength across lineages the way joint multinomial approaches do.

\subsection*{Implications for variant surveillance}

The results presented here suggest a concrete role for DPGR in real-time
variant risk assessment. Genomic surveillance networks (GISAID, GenBank,
national sequencing programs) already generate the only input DPGR requires:
lineage-labeled sequence counts over time. No absolute incidence data,
generation-time estimates, phylogenetic reconstruction, or transmission model
fitting is needed to compute the relative fitness signal.

Three empirical findings support this application. First, the early warning
analysis shows that a sustained DPGR signal ($> 0.005$) preceded variant
dominance by 43 to 65~days (mean 55~days) across all five major SARS-CoV-2
transitions, with 95\% sign accuracy and a mean absolute error of 11.4~days
for time-to-dominance prediction. In a real-time setting, such lead time could
support earlier prioritization of vaccine-strain review, non-pharmaceutical
intervention planning, and healthcare-capacity planning, subject to local data
quality and independent epidemiological assessment. Second, cross-country
concordance shows that the direction of the fitness advantage is globally
consistent in the data analyzed here, so an early DPGR signal detected in one
country may help prioritize monitoring elsewhere, even though the magnitude
varies with local immune landscapes. Third, the detection threshold
of $|\mathrm{DPGR}| = 0.004$ at 80\% power is roughly 10$\times$ below
typical variant-replacement signals ($|\mathrm{DPGR}| \sim 0.04$), so the
method is sensitive enough to detect emerging fitness advantages well before
they produce visible frequency shifts.

In practice, the workflow would be: a surveillance laboratory computes
sliding-window log-ratio regressions from routine sequencing data; a
sustained positive DPGR triggers a variant assessment; and the DPGR
magnitude, combined with the current frequency, yields a quantitative
estimate of the time to variant dominance. Because DPGR is computationally
trivial (milliseconds per comparison) and requires no tuning beyond the
window size, it can be deployed as an automated early-warning layer in
existing surveillance pipelines. The growth-rate bridge developed here makes
the resulting fitness estimates interpretable as $R_t$ contrasts only after a
generation-interval model, baseline growth rate, and relevant surveillance
assumptions have been specified.

\subsection*{Interpreting regression slopes below~1}

Regression slopes below~1.0 appear in all empirical analyses
(Table~\ref{tab:results}), but the SIR simulation establishes that this is
not a failure of the SIR growth-rate mapping: with the true instantaneous~$R_t$,
the expected mapping is recovered (slope~$= 0.99$; Figure~\ref{fig:simulation}).
The attenuation arises entirely from the estimation procedure. As shown in
Table~\ref{tab:slope_correction}, matching the Cori smoothing window to the
DPGR window and substituting real case counts for sequence-count proxies each
independently improve slopes toward~1.0. These are properties of the Cori
estimator, not of the DPGR/$R_t$ relationship.

This has a methodological implication: correlation and sign agreement, which
are invariant to multiplicative scale factors, are the appropriate metrics
for checking consistency of the mapping. Regression slopes are informative about
estimation-pipeline calibration but should not be mistaken for tests of the
underlying theory.

\section{Conclusion}

The mathematical connection follows once DPGR is interpreted as a
pairwise growth-rate difference. In a short-window two-variant SIR model,
\[
\ln(10)\,\mathrm{DPGR}_{i,j} = r_i-r_j
= \gamma_i(R_{t,i}-1)-\gamma_j(R_{t,j}-1).
\]
Hence:
\begin{itemize}
    \item DPGR is best viewed as a \emph{relative transmission fitness}
    statistic.
    \item Proposition~\ref{prop:sir-bridge} shows that, under the short-window
    two-variant SIR approximation, DPGR is linked to variant-specific
    $R_t$ and, in the fully susceptible limit, to variant-specific $R_0$.
    \item Proposition~\ref{prop:abstract-bridge} shows that beyond strict SIR,
    DPGR continues to encode the pairwise growth-rate contrast and therefore
    induces a contrast in reproduction-number space only after a
    generation-interval model and baseline growth rate are specified.
\end{itemize}

Together, Propositions~\ref{prop:sir-bridge} and
\ref{prop:abstract-bridge} summarize the two main levels of interpretation:
a concrete SIR identity and a more general growth-to-reproduction bridge.
More broadly, DPGR, Kimura et al.'s softmax slopes, and the Figgins-Bedford
GARW parameters can be compared on a common pairwise growth-advantage scale
under aligned assumptions about time scale, generation interval, sampling, and
local growth; they should not be treated as interchangeable estimators.

This makes DPGR a natural complement to $R_t$: DPGR is optimized for comparing
variants against each other, whereas $R_t$ is optimized for measuring absolute
transmission intensity over time. Beyond the mapping checks, the evolutionary analyses
show that DPGR captures biologically interpretable signal in these data: it is
approximately transitive at the lineage level, approximately zero for selected
functionally similar sub-lineages, and directionally consistent across
countries.

Beyond its theoretical interest, this connection has practical surveillance
implications. Because DPGR requires only sequence counts and can be computed
in milliseconds, it can serve as an automated early-warning layer in existing
genomic surveillance pipelines. The 43 to 65~day lead time before variant
dominance observed in these retrospective analyses could support earlier
vaccine-strain review, public-health planning, and healthcare-capacity
planning when corroborated by local epidemiological evidence. By grounding
this surveillance tool in formalized growth-rate
relationships that connect population-genetic fitness theory to epidemic
modeling, we provide a clearer theoretical basis and empirical consistency
checks for using DPGR in variant surveillance.

\section*{Appendix: Formal verification in Lean}

The main statements in Propositions~\ref{prop:sir-bridge} and
\ref{prop:abstract-bridge}, together with the important special cases,
were formalized and machine-verified in Lean~4 (v4.29.0) with Mathlib
(v4.29.0). The formalization will be released after formal publication of this
manuscript and is organized as follows:

\begin{itemize}
    \item \texttt{Formalization/Basic.lean}:
    Core definitions (\texttt{localGrowthRate}, \texttt{effectiveRt},
    \texttt{dpgr}, \texttt{basicR0}, \texttt{generationTime}) and seven
    theorems covering:
    \begin{enumerate}
        \item Proposition~\ref{prop:sir-bridge}, part~1: the SIR bridge
        identity $\mathrm{DPGR} = \frac{1}{\ln 10}[\gamma_i(R_{t,i}-1)
        - \gamma_j(R_{t,j}-1)]$;
        \item Proposition~\ref{prop:sir-bridge}, part~2: the $R_0$ bridge
        when $S=N$;
        \item The equal-$\gamma$ simplification
        $\mathrm{DPGR} = \frac{\gamma}{\ln 10}(R_{t,i}-R_{t,j})$
        (\ref{eq:equal_gamma});
        \item The generation-time form
        $\mathrm{DPGR} = \frac{R_{t,i}-R_{t,j}}{T_g\ln 10}$
        (\ref{eq:tg_form}).
    \end{enumerate}

    \item \texttt{Formalization/AbstractBridge.lean}:
    Proposition~\ref{prop:abstract-bridge}: for an arbitrary map $\Phi$
    from growth rate to reproduction number,
    $R_{t,i}-R_{t,j} = \Phi(r_j+\ln 10\cdot\mathrm{DPGR})-\Phi(r_j)$,
    plus the SIR specialization $\Phi(r)=1+r/\gamma$.

    \item \texttt{Formalization/ReLogRatio.lean}:
    The fixed-generation-time $R_e$ log-ratio identity
    $\mathrm{DPGR} = \frac{1}{T_g}\log_{10}(R_{e,i}/R_{e,j})$
    (\ref{eq:dpgr_re_logratio}).
\end{itemize}

In the authors' local version, all proofs compile without \texttt{sorry},
\texttt{admit}, or custom axioms. Verification instructions will be included
with the released formalization.

\section*{Data and Code Availability}

SARS-CoV-2 sequence metadata were obtained from the GISAID EpiCoV
database (\url{https://www.gisaid.org}); access requires registration and
acceptance of the GISAID Data Access Agreement. Influenza sequence data
were obtained from GISAID EpiFlu. UK case incidence data are publicly
available from the UK Health Security Agency. Analysis code and the Lean~4
formalization will be released after formal publication of this manuscript.

\section*{Acknowledgments}

We gratefully acknowledge all data contributors, i.e., the Authors and
their Originating laboratories responsible for obtaining the specimens,
and their Submitting laboratories for generating the genetic sequence and
metadata and sharing via the GISAID Initiative, on which this research is
based.

This work was supported by the National Science Foundation under awards
2525493 and 2200138 (Predictive Intelligence for Pandemic Prevention,
PIPP Phase~I).

Hong Qin disclosed a related patent pending.


\begin{thebibliography}{99}

\bibitem{AnDerHeiden2022}
M. an der Heiden and U. Buchholz,
\emph{Serial interval in households infected with SARS-CoV-2 variant
B.1.1.529 (Omicron) is even shorter compared to Delta},
Epidemiology and Infection, 150: e146, 2022.
\href{https://doi.org/10.1017/S0950268822001248}{doi:10.1017/S0950268822001248};
\url{https://doi.org/10.1017/S0950268822001248}.

\bibitem{Annan2025}
R. Annan, U. Nkonu, P. Hatami, M. J. Pantho, L. Qingge, and H. Qin,
\emph{Predicting variant fitness of SARS-CoV-2 from full viral genome sequences},
Proceedings of the AAAI Symposium Series, 7(1): 428--437, 2025.
\href{https://doi.org/10.1609/aaaiss.v7i1.36915}{doi:10.1609/aaaiss.v7i1.36915};
\url{https://doi.org/10.1609/aaaiss.v7i1.36915}.

\bibitem{Cori2013}
A. Cori, N. M. Ferguson, C. Fraser, and S. Cauchemez,
\emph{A new framework and software to estimate time-varying reproduction
numbers during epidemics},
American Journal of Epidemiology, 178(9): 1505--1512, 2013.
\href{https://doi.org/10.1093/aje/kwt133}{doi:10.1093/aje/kwt133};
\url{https://doi.org/10.1093/aje/kwt133}.

\bibitem{FigginsBedford2025}
M. D. Figgins and T. Bedford,
\emph{Inferring variant-specific effective reproduction numbers from combined
case and sequencing data},
eLife reviewed preprint, 2025.
\href{https://doi.org/10.7554/eLife.104802.1}{doi:10.7554/eLife.104802.1};
\url{https://doi.org/10.7554/eLife.104802.1}.

\bibitem{Hatami2026}
P. Hatami, R. Annan, L. Miranda, J. Gorman, M. Xie, L. Qingge, and H. Qin,
\emph{Explainable convolutional neural network model provides an alternative genome-wide association perspective on mutations in SARS-CoV-2},
Scientific Reports, 2026.
\href{https://doi.org/10.1038/s41598-026-53625-x}{doi:10.1038/s41598-026-53625-x};
\url{https://doi.org/10.1038/s41598-026-53625-x}.

\bibitem{KauffmanLevin1987}
S. A. Kauffman and S. Levin,
\emph{Towards a general theory of adaptive walks on rugged landscapes},
Journal of Theoretical Biology, 128(1): 11--45, 1987.
\href{https://doi.org/10.1016/S0022-5193(87)80029-2}{doi:10.1016/S0022-5193(87)80029-2};
\url{https://doi.org/10.1016/S0022-5193(87)80029-2}.

\bibitem{Kimura2022Cell}
I. Kimura, D. Yamasoba, T. Tamura, et al.,
\emph{Virological characteristics of the SARS-CoV-2 Omicron BA.2 subvariants,
including BA.4 and BA.5},
Cell, 185(21): 3992--4007.e16, 2022.
\href{https://doi.org/10.1016/j.cell.2022.09.018}{doi:10.1016/j.cell.2022.09.018};
\url{https://doi.org/10.1016/j.cell.2022.09.018}.

\bibitem{MKimura1962}
M. Kimura,
\emph{On the probability of fixation of mutant genes in a population},
Genetics, 47(6): 713--719, 1962.
\href{https://doi.org/10.1093/genetics/47.6.713}{doi:10.1093/genetics/47.6.713};
\url{https://doi.org/10.1093/genetics/47.6.713}.

\bibitem{MKimura1968}
M. Kimura,
\emph{Evolutionary rate at the molecular level},
Nature, 217: 624--626, 1968.
\href{https://doi.org/10.1038/217624a0}{doi:10.1038/217624a0};
\url{https://doi.org/10.1038/217624a0}.

\bibitem{Obermeyer2022}
F. Obermeyer, M. Jankowiak, N. Barkas, et al.,
\emph{Analysis of 6.4 million SARS-CoV-2 genomes identifies mutations associated with fitness},
Science, 376(6599): 1327--1332, 2022.
\href{https://doi.org/10.1126/science.abm1208}{doi:10.1126/science.abm1208};
\url{https://doi.org/10.1126/science.abm1208}.

\bibitem{Pantho2025}
M. J. Pantho, R. Annan, L. A. Bauder, S. Huang, L. Qingge, and H. Qin,
\emph{A data-driven sliding-window pairwise comparative approach for the
estimation of transmission fitness of SARS-CoV-2 variants and construction of
the evolution fitness landscape},
Quantitative Biology, 13(4): e70003, 2025.
\href{https://doi.org/10.1002/qub2.70003}{doi:10.1002/qub2.70003};
\url{https://doi.org/10.1002/qub2.70003}.

\bibitem{ParkPNAS2023}
S. W. Park, K. Sun, S. Abbott, R. Sender, et al.,
\emph{Inferring the differences in incubation-period and
generation-interval distributions of the Delta and Omicron variants of
SARS-CoV-2},
Proceedings of the National Academy of Sciences, 120(22): e2221887120, 2023.
\href{https://doi.org/10.1073/pnas.2221887120}{doi:10.1073/pnas.2221887120};
\url{https://doi.org/10.1073/pnas.2221887120}.

\bibitem{Uddin2025}
J. M. I. Uddin, M. J. Pantho, and H. Qin,
\emph{Quantifying influenza strain dominance: a differential population growth rate analysis across regions and seasons},
2025 IEEE International Conference on Data Mining Workshops (ICDMW), 1--10, 2025.
\href{https://doi.org/10.1109/icdmw69685.2025.00006}{doi:10.1109/icdmw69685.2025.00006};
\url{https://doi.org/10.1109/icdmw69685.2025.00006}.

\bibitem{WallingaLipsitch2007}
J. Wallinga and M. Lipsitch,
\emph{How generation intervals shape the relationship between growth rates and
reproductive numbers},
Proceedings of the Royal Society B, 274(1609): 599--604, 2007.
\href{https://doi.org/10.1098/rspb.2006.3754}{doi:10.1098/rspb.2006.3754};
\url{https://doi.org/10.1098/rspb.2006.3754}.

\end{thebibliography}
\end{document}